\documentclass[sigconf]{acmart}
\AtBeginDocument{%
  \providecommand\BibTeX{{%
    \normalfont B\kern-0.5em{\scshape i\kern-0.25em b}\kern-0.8em\TeX}}}
\usepackage{amsmath}
\usepackage{footmisc}

\begin{document}

\title{Role of Attentive History Selection in Conversational Information Seeking}

% \author{Somil Gupta}
% \email{somilgupta@umass.edu}
% \author{Neeraj Sharma}
% \email{neerajsharma@umass.edu}
% \affiliation{%
%   \institution{University of Massachusetts Amherst}
%   \streetaddress{}
%   \postcode{01002}
% }

\author{Somil Gupta}
\email{somilgupta@umass.edu}
\affiliation{%
  \institution{University of Massachusetts Amherst}
  \streetaddress{}
  \postcode{01002}
}

\authornote{Both authors contributed equally to this research.}
\orcid{1234-5678-9012}
\author{Neeraj Sharma}
\authornotemark[1]
\email{neerajsharma@umass.edu}
\affiliation{%
  \institution{University of Massachusetts Amherst}
  \streetaddress{}
  \postcode{01002}
}

% \affiliation{%

% \affiliation{%

% \affiliation{%


\begin{abstract}
The rise of intelligent assistant systems like Siri and Alexa have led to the emergence of Conversational Search, a research track of Information Retrieval (IR) that involves interactive and iterative information-seeking user-system dialog. Recently released OR-QuAC and TCAsT19 datasets narrow their research focus on the retrieval aspect of conversational search i.e. fetching the relevant documents (passages) from a large collection using the conversational search history. Currently proposed models for these datasets incorporate history in retrieval by appending the last $N$ turns to the current question before encoding. We propose to use another history selection approach that dynamically selects and weighs history turns using the attention mechanism for question embedding. The novelty of our approach lies in experimenting with soft attention-based history selection approach in open-retrieval setting.
 
\end{abstract}




\maketitle
\section{Introduction} 
\par Conversational Search is a long-standing goal of the Information Retrieval (IR) community that envisions IR systems to be able to ascertain and satisfy user information need iteratively and interactively. The emergence and subsequent popularity of intelligent assistant systems like Siri, Alexa, AliMe, Cortana, and Google Assistant, has further fuelled research and investment in this domain. However, conversational search by itself envelopes numerous settings, each with their problems to tackle, so the researchers tend to narrow their research focus on specific sub-problems. One such setting can be "System Ask, User Respond" \cite{user_ask_sys_resp} where the system can ask user proactively to clarify their information need. 
\par A complementary setting involves the system responding with the answer without any clarification questions. Conversational Question Answering (ConvQA) \cite{DBLP:conf/sigir/Qu0QCZI19} is an example of this setting where the system answers user questions using the given evidence. CoQA \cite{DBLP:journals/tacl/ReddyCM19} and QuAC \cite{choi2018quac} are common ConvQA datasets. However, these datasets primarily focus on answer formulation given the evidence and ignore the retrieval aspect of the problem. Recently released open-retrieval datasets OR-QuAC \cite{Qu_2020} and TCAsT19 \cite{DBLP:journals/corr/abs-2003-13624} focus on the retrieval aspect of the conversational information-seeking problem. The focus of our research will be on these \textit{open-retrieval conversational question answering models} \cite{Qu_2020}. 
\par History Modelling and Selection is an important aspect of conversational search which dictates how the previous conversational turns would be incorporated into the model while the model reasons on the current query. Different modeling and selection approaches are used across convQA models \cite{gupta2020conversational}. The current open-retrieval models append the last $N$ turns of the conversation with the current question and expect the model to derive context. This assumes that immediate turns have all the important contextual information for the current question. However, conversational dialogs may have \textit{topic shifts} (the current question is not immediately relevant
to something previously discussed), or
\textit{topic returns} (the current question is asking about a
previously shifted topic)  \cite{dataset_comparison}. Instead we explore another  history selection technique suggested by Qu et. al.\cite{Qu_2019} for ConvQA setting that employs \textit{attention mechanism} to find the relative importance of each turn to the current question. Such a technique does not make any assumption and lets the model attend over all the history turns to decide which ones to include. Since history attention mechanism improved the results in ConvQA setting, it may be worth experimenting within the retrieval setting to identify the relative weightage of a history question while preparing the combined question embedding.  

\section{Problem Statement}
 The task is to retrieve a ranked list of passages in response to a question by using previous questions in conversation as context. 
Formally this problem can be defined as follows: Given a dialog $D$ with conversation turns or questions, $Q = \{q_1, q_2, ..., q_n\}$  and a collection $C$, the task is to retrieve a ranked list of $R$ passages $\{P\}_{r=1}^R$ from $C$, based on their relevance to each question $q_k \in Q$ using preceding context $\{Q\}_{i=1}^{N-1}$ weighted by $\{\alpha\}_{i=1}^{N-1}$ where $\alpha$ is the attention weight computed for that turn. The soft attention $\alpha_i$ in this case would denote the relative co-reference of the current question $q_k$ to the history question $Q_i$. We ignore the reader and the ranker subsections in OR-QuAC implementation \cite{Qu_2020} as our primal focus is to modify history selection for open-retrieval.

\section{Related Work}
\par \textbf{Conversational Information Seeking:} In open retrieval settings, \cite{lee2019latent}, \cite{DBLP:conf/iclr/DasDZM19}, and \cite{DBLP:conf/emnlp/KarpukhinOMLWEC20} adopted a dual-encoder architecture to construct a learnable retriever and demonstrate that their methods are scalable to large collections, but these works are limited to single-turn QA settings. In \cite{Qu_2020}, the authors illustrate the approaches for open-retrieval question-answer setting for multiple conversational turns. In an open-retrieval setting, the retrieval process is open in terms of retrieving from a large collection instead of re-ranking a small number of passages in a closed set. Their model elaborately defines a retriever, a re-ranker, and a reader based on the Transformer model \cite{DBLP:conf/nips/VaswaniSPUJGKP17}. History modeling is done by concatenating previous N history questions to the current question. Our approach would use their model as our underlying implementation, however, we would disable the reader and ranker modules (only enable retrieval), and use attention mechanism for history selection in the retriever module.

\par \textbf{History selection:} Our attention based history selection approach is inspired by the work by Qu et al. \cite{Qu_2019} who employ soft attention mechanism to select and weigh history turns based on how helpful they are in answering the current question in the ConvQA setting. However, while they tried to use attentive history selection in ConvQA (answer formulation), we will try to employ it for retrieval to form a single query embedding for passages.

\begin{figure}[h!]
     \centering
    \makebox[\columnwidth]{\includegraphics[scale=0.38]{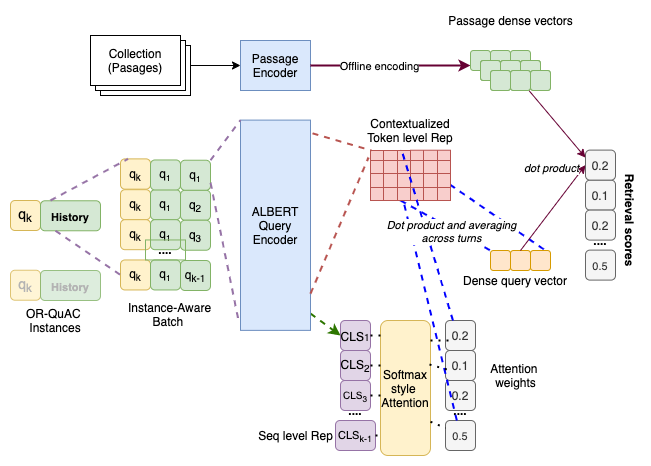}}
  \caption{A high-level illustration of Fine-grained History Attentive Retriever (HAR). Each OR-QuAC instance is converted into a instance-aware sub-batch of size equal to the number of turns in the instance, where each row of this batch contains the tokens of the current question $q_k$, history turn $q_{i<k}$ and CANARD rewritten first question $q_1$. This batch is passed through an ALBERT query encoder to create sequence level contextualized representations ($CLS_{i}$) and token-level representations for each turn. Sequence-level representations of all history turns in the batch are used to generate soft attention weights $\alpha_{i<k}$ using softmax-style attention. Weighted averaging at token-level across the batch on contextualized token representation
  followed by subsequent averaging produces the dense question vector. In a separate workflow, each passage in the collection $C$ is encoded offline to produce dense vectors. A similarity function (in our case, dot product) can finally be used to obtain relevance scores per paragraph for the given question $q_k$ and history $q_{i=1}^{k-1}$. }
    \label{fig:ham-retrieval}
\end{figure}
\section{Approach}
We use the same underlying architecture and training/evaluation process as in OR-ConvQA \cite{Qu_2020}, using its model as our baseline and its code repository\footnote{\label{orconvqa}\url{https://github.com/prdwb/orconvqa-release}} as our starting point. We add \textbf{three} major modifications to the baseline model implementation. \label{label:data-preprocessing}
\begin{enumerate}
\itemsep1em
    \item We modify the baseline retriever to use \textit{soft attention mechanism} for history modelling into the current query \textit{during concurrent learning}. Please note that \textit{no changes are made during retriever pre-training} as it is trained on CANARD\cite{elgohary-etal-2019-unpack} rewrites instead of QuAC\cite{choi2018quac} questions and therefore has complete context in the query itself. We call this modified retriever model as \textbf{History Attentive Retriever (HAR)}. 
    \item We need to compute attention using the contextualized output of our query encoder for each history turn, requiring modification of input batches to contain all history turns into a single batch. We, therefore, make changes to batch preprocessing for retriever model and name this batching process as \textbf{instance-aware batching} similar to HAM\cite{Qu_2019}.
    \item Since the focus of our experiment in on retrieval process in concurrent learning, we modify our model checkpoint selection process during evaluation phase to\textit{ use retriever-recall instead of reader-F1 score as the selection metric}.  
\end{enumerate} 
A high-level description (and illustration) of the modified retrieval model is provided in Figure \ref{fig:ham-retrieval}. Subsequent sub-sections discuss some important aspects of our approach. 
\subsection{Instance-aware batch preprocessing}
\par The OR-QuAC retriever instance consists of the current question, the history turns (questions and answers for those questions). For each instance, OR-QuAC creates a sequence of current question prepended by history questions \textit{for a fixed-window size}, and after tokenization, the input batch to the baseline retriever has dimensions $(N, M, B)$ where $N$ is the number of retriever batch size per GPU, $M$ is the max ALBERT sequence length and $B$ is token embedding size.  
\par However, HAR translates each OR-QuAC instance into a collection of sequences for \textit{all history turns} where each sequence contains the first question (for complete context), turn question and the current question. Each of these sequences is an input to ALBERT and their contextualized output is recombined to generate the query vector. In order to perform this computation together, we need to include all these sequences related to a single instance in a single batch. This batch is called \textbf{instance-aware batch}. A typical batch in baseline model would translate to  $(N, I, M, B)$ where $I$ additionally is the max number of history turns in the instance and $N$ is the number of retriever batch size per GPU. For the sake of simplicity, we do not combine examples from multiple instances in the same retriever GPU batch as implemented in HAM\cite{Qu_2019}. 
\subsection{History Attentive Retriever (HAR)}
\label{label:har}
We use ALBERT \cite{albert} for contextualized question encoding and ALBERT tokenizer for tokenization. Let $N$ be the max number of history turns, $M$ be the max number of tokens in a question, $q_{k}$ be the current question and $q_{i<k}$ be the $i^{th}$ history question in history $\{q\}_{i=1}^k$. Then, 
\begin{align*}
    \mathcal{I}_i &= \text{Tokenizer}(\text{[CLS] }q_1 \text{ [SEP] } q_i \text{ [SEP] } q_k)\\
     \mathbf{G_i} &= 
    \text{ALBERT}(\mathcal{I}_i, \text{PosSeg}[i])
\end{align*}
     where, 
\begin{itemize}
     \item $\mathcal{I}_i \in \mathbb{R}^{(3M+3) \times b}$ is the input embeddings for ALBERT for turn $i$ with $b$ as the embedding size of the tokenizer. Here,  $\mathcal{I}$ is the instance-aware batch. 
     \item PosSeg$[i] \in \mathbb{R}^{(3M+3) \times b}$ is the \textbf{positional segment embedding} for turn $i$ (discussed in \ref{pos-seg}). 
    \item $\mathbf{G_i}\in \mathbb{R}^{(3M+3) \times b}$ is the contextualized output of ALBERT for turn $i$.
    \item $\mathbf{s_i} = \mathbf{G_i}[CLS]$ is the contextualized sequential representation corresponding to CLS token for turn $i$ where $\mathbf{s_i} \in \mathbb{R}^{b}$.
    \item $\mathbf{T_i}[m] = \mathbf{G_i}^{q_k}[m]$ is the contextualized token representation corresponding to the $m^{th}$ token of $q_k$ w.r.t. $q_i$ and $q_1$, where $\mathbf{T_i} \in \mathbb{R}^{M \times b}$ for turn $i$.
\end{itemize}

\par Now, taking $\mathbf{d} \in \mathbb{R}^{b}$ as the attention learnable parameter, 
$$\alpha_i  = \frac{\exp{\mathbf{d}^T\cdot\mathbf{s_i}}}{\sum_{i'=1}^N \exp{\mathbf{d}^T\cdot\mathbf{s_{i'}}}} \text{, where } \alpha_i \in [0,1]$$ where $\alpha_i$ is the attention weight of turn $i$ over $q_k$.
\par Once we have the attention weights over the entire history turn, there are \textbf{two} ways to generate dense query vector $\hat{q_k} \in \mathbb{R}^{b}$.
\begin{itemize}
    \item \textbf{Fine-grained history selection} as depicted in Figure \ref{fig:ham-retrieval}, we apply soft-attention at token level and average all the final token representations, i.e.
    \begin{align*}
    \hat{Q_{k}} &= \sum_{i=1}^N  \alpha_i \mathbf{T}_{i}\text{  ,where  } \hat{Q}_{k} \in \mathbb{R}^{M \times b}\\
    \hat{q_k} &=  \frac{1}{M}\sum_{i=1}^M \hat{Q}_{k,m}
    \end{align*}
  \item \textbf{Coarse-grained history selection:} we can apply soft-attention at sequence level using CLS representations of each turn, i.e.
    $$\hat{q_{k}} = \sum_{i=1}^N  \alpha_i \mathbf{s}_{i}$$
\end{itemize} 
Finally, retriever scores can be computed using the passage dense vectors $\{p_j\}_{j=1}^{J}$ where $J$ is the total number of candidate passages obtained using Faiss index retriever as in the baseline model\cite{Qu_2020}. $$\text{retriever score}(\hat{q}_k, p_j) = \hat{q}_k\cdot p_j$$
\par [Please note that the embedding dimension $b$ is generally different from the BERT hidden size $h$, due to which we have to use input/output projection linear layers for $\mathbb{R}^{M \times h} \implies \mathbb{R}^{M \times b}$, however, we have ignored it for the sake of simplicity.] 
\subsection{Positional Segment Embedding}
\label{pos-seg}
\par Qu et. al. \cite{Qu_2019} found that encoding relative position of the history turn from the current turn during history modelling improves model performance for ConvQA. They introduce an additional positional history answer embedding to augment BERT input embeddings.

\par In order to test this hypothesis for open-retrieval, we encode the relative turn information in segment input embeddings that are used by ALBERT to differentiate different segments in a sequence \cite{albert}. \textit{We annotate tokens of current question $q_k$ with token type identifier $0$ and tokens of (first question $q_0$ + turn history question  $q_i$) with token type identifier $k-i$}.  The token type vocabulary size in ALBERT configurations is set to max history turns $N$. This helps to map history turn input segments to different segment embedding tokens.

\section{Experiment}
\label{experiment}
\subsection{Datasets}
     \par We use OR-QuAC \cite{Qu_2020} as our primary dataset for experiments. OR-QuAC enhances QuAC by adapting it to an open retrieval setting. It is an aggregation of three existing datasets: (1) the QuAC dataset \cite{choi2018quac} that offers information-seeking conversations, (2) the CANARD dataset \cite{elgohary-etal-2019-unpack} that consists of context-independent rewrites of QuAC questions, and (3) the Wikipedia corpus that serves as the knowledge source of answering questions. Data statistics of the OR-QuAC dataset are mentioned in table \ref{tab:or-quac}. The dataset can be downloaded from the CIIR page\footnote{\label{dataset}\url{https://ciir.cs.umass.edu/downloads/ORConvQA/}}.  

\begin{table}[!tbh]
\caption{Data Statistics of OR-QuAC dataset \cite{Qu_2020}}
\label{tab:or-quac}
\centering
\begin{tabular}{llll}
\toprule
Items                                     & Train      & Dev        & Test       \\
\midrule
\# Dialogs                                & 4383       & 490        & 771        \\
\# Questions                              & 31526      & 3430       & 5571       \\
\# Avg. Question Tokens                   & 6.7        & 6.6        & 6.7        \\
\# Avg. Answer Tokens                     & 12.5       & 12.6       & 12.2       \\
\# Avg. Question/Dialog                   & 7.2        & 7.0        & 7.2        \\
\begin{tabular}[c]{@{}l@{}}\# Min/Avg/Med/Max\\  History Turns/Question
\end{tabular} & 0/3.4/3/11 & 0/3.3/3/11 & 0/3.4/3/11 \\
 \bottomrule
\end{tabular}
\end{table}

 \subsection{Baseline}
\label{baselines}
We compare our results against the OR-ConvQA system defined in the OR-QuAC paper \cite{Qu_2020}. Model\footnote{\label{baseline}https://github.com/prdwb/orconvqa-release} was run using the hyperparameters as defined in Table \ref{tab:setup}. Additionally, history window size was set to $6$ and only history questions were prepended alongwith the first question. Checkpoint at $90000$ training steps was selected as the best model using Reader-F1 metric during evaluation and corresponding test results are reported in Table \ref{tab:or-quac-result}. 

\subsection{Implementation Details}
We implemented our approach for HAR as defined in section \ref{label:data-preprocessing} using OR-QuAC repo\footref{baseline} as our underlying implementation. The retriever checkpoint for pre-training and the passage representations were reused from OR-ConvQA repo\footref{dataset}. Both the fine-grained and coarse-grained versions of HAR were implemented and run with positional Segment Embeddings enabled, using hyper-parameters as defined in Table \ref{tab:setup} (same as the baseline for comparison). Models
are trained with 2 NVIDIA TITAN X GPUs- 1 for training and another for Faiss\footnote{\url{https://github.com/facebookresearch/faiss}} indexing. Retriever recall is used as model selection metric during evaluation as the best model checkpoint is used for testing. Our implementation is located on Github \footnote{\url{https://github.com/somiltg/orconvqa-release}}.

\begin{table}[tbh!]
\caption{Hyperparameter values for HAR model's concurrent learning and evaluation. }
\centering
\begin{tabular}{ll}
\toprule
\textbf{Hyperparameter}              & \textbf{Value} \\ 
\midrule
Max (seq, ques, passage, ans) length & 512, 125, 384, 40    \\ 
Train, eval batch sizes per GPU         & 1,1           \\ 
Learning rate    & 5e-5      \\
Number of epochs                     & 3.0            \\
Top k results for (retriever, reader)                    & 100,5            \\
Max History Turns & 11\\
Max training iterations & 90000\\
 \bottomrule
\end{tabular}%

\label{tab:setup}
\end{table}

\subsection{Evaluation}
We used Mean Reciprocal Rank \textbf{(MRR)} and \textbf{Recall} to evaluate the retrieval performance for the proposed history attention retriever (HAR). The reciprocal rank of a query is the inverse of the rank of the first positive passage in the retrieved passages. \textit{MRR} is the mean of the reciprocal ranks of all queries. \textit{Recall} is the fraction of the total amount of relevant passages that are retrieved. Table \ref{baselines} provides our results on OR-QuAC test data for best performing model checkpoints\footnote{\label{checkpoints}\url{https://umass.box.com/s/uya1t53h3qfro0x8i9k4joivcmsy08fa}} of HAR variations and baseline implementations. 
\begin{table}[!tbh]
\centering
\caption{Test Results of History Attentive Retriever(HAR) implementations along with the baseline.}
\label{tab:or-quac-result}
\begin{tabular}{lll}
\toprule
Model & MRR                                     &  Recall  \\
\midrule
OR-ConvQA baseline \cite{Qu_2020} & 0.2166                                &   0.3045     \\
 \bottomrule
 HAR w/ fine-grained attention & 0.1995 &  0.2742 \\
 HAR w/ coarse-grained attention & 0.1966 & 0.2812 \\
  \bottomrule
\end{tabular}
\end{table}
\section{Analysis}
\subsection{Result Analysis}
We present an analysis of the test results presented in Table \ref{tab:or-quac-result}.
\begin{itemize}
\itemsep1em
\item Our results show that \textbf{HAR with coarse-grained attention performs better on Recall than fine-grained attention}. A possible hypothesis for this can be that ALBERT CLS token is more appropriate for capturing abridged latent information about the complete query compared to token-based representation. Fine-grained representation may additionally be capturing low-level unessential details of the query which may not be relevant for query passage interaction. Also, we use \textit{mean} to obtain a single dense query vector from token-level query representation (obtained via soft attention scores based weighted averaging of ALBERT outputs across conversation turns). 
However, mean might be phasing out important latent information or amplifying unimportant one in its effort to grant equal weight to each token. A more elaborate aggregation mechanism like a linear layer might be more suitable. Our conclusion about fine-grained performing better is also corroborated by OR-QuAC\cite{Qu_2020} results which also uses CLS based representation for query vector. 
\par \textbf{While there is a noticeable difference in recall scores, the results on MRR are almost the same on both the variants.} This suggests that the topmost relevant document position in the results is very much same in both variants. 
\item \textbf{Our HAR approach is performing comparatively decently to the baseline but is not able to beat the performance of the baseline with any of the variants}. One possible reasoning may the inappropriate use of an aggregation function for fine-grained attention which may have boosted the performance. Another reason may be the use of pretrained retriever model used by the baseline instead of running our own using attention approach, which might be beneficial for our approach. Finally, the reader model may not be in complete resonance with this retriever approach and may need to be differently fine-tuned. However, we have bucketed both the retriever and reader models together (and train them with the same back-propogation) which might cause variations in one (say reader) to overtly penalize the other(say retriever). 
\end{itemize}
\subsection{Ablation Analysis}
We conduct ablation studies on HAR to assess the importance of each sub-component in our approach. Table \ref{tab:ablation-studies} provides results on different HAR variations, ablating positional segment embedding and soft-attention ($\alpha=1$) for both fine- and coarse-grained models. 
\begin{table}[!tbh]
\centering
\caption{Ablation Studies for HAR}
\label{tab:ablation-studies}

\begin{tabular}{lll}
\toprule
Model & MRR                                     &  Recall  \\
\toprule
\textbf{HAR w/ fine-grained attention} & 0.1995 &  0.2742 \\
\bottomrule
w/o PosSeg & 0.1902 & 0.2635 \\
\bottomrule
\textbf{HAR w/ coarse-grained attention} & 0.1966 & 0.2812 \\
\bottomrule
w/o PosSeg & 0.1960 & 0.2819 \\
w/o Soft Attention ($\alpha=1$) & 0.1948 & 0.2773 \\ 
\bottomrule
\end{tabular}
\end{table}
\begin{itemize}
\itemsep1em
\item \textbf{Addition of positional segment embedding improves performance on both metrics for fine-grained HAR but not for coarse-grained attention}. A possible reasoning for this can be that fine-grained attention works at token level where encoding positional turn information may help differentiate tokens across conversational turns during soft-attention weight computation or weighted averaging. For fine-grained attention, this difference amplifies as it is computed at token level, while for coarse-grained attention, CLS token concerns with capturing overall sentence representation and therefore token level differentiation may not help.
\item \textbf{Soft-attention over history turns improves performance on both metrics.} This is clearly shown by the results of fine-grained HAR w/o soft Attention which sets soft-attention scores to 1 for all conversational turns. This justifies our claim that different history turns have different contribution to the current question and can improve retrieval. One can refer to the analysis in HAM\cite{Qu_2019} that provides a heat map of topic return, topic return and drill down - important phenomena in conversations that are effectively captured by soft-attention based history modelling. 
\end{itemize}

\section{Conclusion}
In this work, we proposed a history attention retriever mechanism for Open Retrieval question answer settings. We used a concept introduced in \cite{Qu_2019} called history attention mechanism to calculate the dense representation of query using the history queries asked in the same session. We show that our model can effectively capture the
utility of history turns. We conducted extensive experimental evaluations to demonstrate
the effectiveness of history attention retriever with positional segment embeddings. A possible future work in this direction may include exploring performance with history attention based reader model, modifying retriever pretraining to also train on conversational history and/or trying different aggregation approaches with fine-grained attention.


\bibliographystyle{ACM-Reference-Format}
\bibliography{sample-base}

\end{document}